\documentclass[preprint]{article}
\usepackage{graphicx}% Include Figure files
\usepackage{dcolumn}% Align table columns on decimal point
\usepackage{bm}% bold math
\usepackage{amsmath}
\usepackage{amssymb}
\usepackage{multirow}
\usepackage{caption}
\usepackage{epstopdf}
\usepackage{hyperref}

\begin{document}
{\setlength{\oddsidemargin}{1.2in}
\setlength{\evensidemargin}{1.2in} } \baselineskip 0.55cm
\begin{center}
{\LARGE {\bf Durgapal-Fuloria Bose-Einstein condensate stars within  $ f(R,T) $ gravity theory }}
\end{center}
\date{\today}
\begin{center}
  Meghanil Sinha*, S. Surendra Singh \\
Department of Mathematics, National Institute of Technology Manipur,\\
Imphal-795004,India\\
Email:{ meghanil1729@gmail.com, ssuren.mu@gmail.com}\\
 \end{center}
 
\textbf{Abstract}: This manuscript studies the Bose-Einstein condensate (BEC) stars in the light of $ f(R,T) $ gravity here with  Durgapal-Fuloria (DP) metric ansatz. The function under this study features as $ f(R,T) = R + 2\eta T $, where $ \eta $ represents the coupling constant. With the help of it, we have formulated a stellar model describing the isotropic matter here within. Our analysis covers energy conditions, equation of state (EoS) parameter and gradients of the energy-momentum tensor components for a valid BEC stellar framework within $ f(R,T) $ gravitational theory with satisfactory results. The model's stability has been validated via multiple stability criteria viz., the velocity of sound, study of adiabetic index and surface redshift where all are found to be lying within the acceptable range for our stellar model. Thus in all the cases we have found our model to be stable and realistic. From the graphical representations the impact of the coupling constant and the parameter of the DP metric potential are clearly visible. Thus we can state that with all the above-mentioned features we have introduced new stellar solutions for BEC stars with enhanced precise results in this modified gravity.\\


\section{Introduction}\label{sec1} 

\hspace{0.7cm}Recent evidence suggests that Universe is expanding at an accelerating rate. The late-time acceleration has been a focus of increasing research efforts, where findings revealed the mysterious dark energy (DE)'s role in it \cite{G2,G3}. Studies indicate the Universe's distribution to be roughly of ordinary matter, dark matter (DM) and DE \cite{G9,G10}. General relativity (GR) doesn't adequately address these certain gravitational phenomena. Thus alternative approaches were pursued to understand these. Extending GR, modified gravity theories emerged which offer new insights for such cases, which could account for the cosmic acceleration and for the inflation. Revising the Einstein-Hilbert (EH) action with new curvature function can offer the desired path for it. As a result, various modified gravities were studied in the recent studies \cite{G12,G14,G17,G19}. Among them, the gravity which is defined by the arbitrary function of the Ricci scalar $(R)$ is the $f(R)$ gravity, which is presented in detail in the foundational studies of \cite{Q1,Q2}. Geometric insights into inflation and DE are offered by this modified theory \cite{Q3,Q6,Q7}. Modifying $R$ leads to the $f(R,T)$ gravity theory. This theory was developed by Harko and his team \cite{RT1}. This gravitational theory holds significant promise in cosmology and astrophysics. It incorporates matter's energy momentum trace into the EH action. This $ f(R,T) $ gravity generalizes $ f(R) $ gravity, accounting for the quantum effects via the inclusion of the trace of the energy momentum $ T $. Although the study of Universe's nature continues to be an intriguing puzzle, the celestial objects such as compact stars, black holes which offer insights into the hidden secrets of the cosmos, and also might help solving those puzzles. This theory provided Tolman-Oppenheimer-Volkoff (TOV) equation for stable compact objects \cite{RT2}. Research flourished on the applications of this theory to investigate various events \cite{Q9,Q10,Q11}. Studies utilized this $ f(R,T) $ theory to analyze stability in spherically symmetric models \cite{G22}. Spherical symmetry's instability was analyzed through various perturbation schemes \cite{G23,G24,G25}. $f(R,T)$ gravity's implications for stellar models were investigated in recent studies \cite{Q18,Q19,Q20}. A comparative study of $ f(R,T) $ gravity and GR was also conducted in \cite{A1}. Additional investigations were also conducted in multiple fronts in this context \cite{A2,A3}. Structural scalars have revealed compact object evolution patterns in $ f(R,T) $ gravity which we can see in \cite{RT3}. Irregularity factors for self-gravitating stars evolving with imperfect fluid were also explored in this gravitational framework \cite{RT4}. Researchers have delved into the study of collapsed structural frameworks within this theory, where different $ f(R,T) $ models were analyzed for their effect on these objects' properties, contrasting with the GR findings \cite{RT5,RT6,RT7,RT8}. \\
The Large Hadron Collider's Higgs Boson findings backs scalar field existence \cite{R1,R2}. Scalar fields can coalesce into boson stars through gravitational interaction \cite{R3,R5,R6}. Low-temperature scalar fields can lead to BECs. Bose and Einstein's hypothesized BEC state was first achieved using ultra-cold rubidium atoms confined magnetically \cite{R7,R8}. From the experimental outcomes, we can see trapped dilute BEC gases show phase transition to coherence \cite{R9,R10,R11}. Boson stars, made of coherent BECs are possible on cosmic scales. Lab experiments with lasers can replicate gravity bound BEC behavior \cite{R12}. BEC stars formed by self-gravity have sparked significant curiosity among the researchers \cite{R13,R14,R16,R18,R20}. Gravitationally confined BECs could considerably comprise a significant portion of the Universe's DM. Extensive research has been done on BECs in astrophysical and cosmological contexts \cite{R22,R24,R31,R,S}. Studies also have focussed on the neutron stars, which may harbor Bose condensates, coexisting with fermions \cite{R33}. Neutron star superfluidity via Cooper pairs allows for the BEC matter which we can see in \cite{R34}. Here it can be seen that temperature introduces non-condensate excitations alongside pure condensate in BEC stellar matter. The idea of rotating BEC stars has been explored in multiple studies. Findings indicate non-relativistic rotation in boson stars is viable smaller objects \cite{RTT1}. Cold Bose stars were also modeled using Gross-Pitaevskii and Poisson equations \cite{RTT2}. Studies solved Lane-Emden equation for polytropic BEC stars. The non-relativistic approach revealed slow rotation's influence on the BEC halos \cite{RTT3}. It was shown that the relativistic effects might constrain BEC star stability \cite{RTT4}.\\
Research was further extended on BEC stars which included finite temperature stellar study, covering dark matter and compact objects \cite{RTT5,RTT6,RTT7}. Later boson stars were being developed including chemical and temperature effects \cite{RTT8,S1}. Research has considered finite-temperature effects on condensates which are evident from \cite{RTT10,RTT11,RTT12}. A two fluid model of BEC at finite temperature which includes condensates and thermal components has been analyzed through \cite{RTT6}. An analytical EoS for BEC at finite temperature was formulated. On this backdrop, BEC dark matter's cosmological evolution was explored in a flat Friedmann-Robertson-Walker (FRW) geometry. Global properties of static, finite-temperature BEC stars were also examined by the researchers \cite{RTT7}. Studies focussed on the magnetic field's impact on relativistic stars at finite temperature \cite{RTT13}. Thus in the light of the inspiration drawn from these previous works, we aim to explore the zero temperature BEC stellar framework in $ f(R,T) $ modified gravitational theory. In this gravitational theory sometimes finding the exact solution for the field equations poses difficulties to overcome with. Prescribing specific metric functions or electric fields aids in solving these field equations in this particular theory. Thus we seek the help of specific metric function which would help in understanding the stellar matter confinement within. Here DP metric serves as the best and unique choice for the foundation of interior spacetime description regulating the matter description within. It is well known for its smooth nature devoid of any type of singularities and which also gives well behaved state variable profiles in case of the compact stellar configurations.\\
Hence in this interesting and unique physical scenario, we have presented in this paper the DP BEC star with all its physically viable features in  $ f(R,T) $ modified gravitational theory. The manuscript proceeds as follows : Section (\ref{sec1}) presents a concise introduction to $ f(R,T) $ gravity and BEC stellar concept. The general formulation of $ f(R,T) $  gravity along with DP metric function is outlined in section (\ref{sec2}). Section (\ref{sec3}) covers the BEC EoS at finite temperature and stellar structure implications including the energy conditions and gradients of the tensor components. BEC star stability is presented in section (\ref{sec4}) via different approaches and finally in section (\ref{sec5}) we conclude our study.\\

\section{Mathematical framework}\label{sec2}

\hspace{0.7cm}The action for the $f(R,T)$ theory is specified by \\
\begin{equation}\label{1}
\verb"S" = \frac{1}{16 \pi} \int d^{4}x f(R,T) \sqrt{-g} + \int d^{4}x \hat{L}_{M} \sqrt{-g}
\end{equation}\\
with $ f(R,T) $ as function of $ R $ and $ T $, $ \hat{L}_{M} $ represents the matter Lagrangian with $ g $ as the metric determinant. We set $ G = c = 1 $ in our calculations throughout. Variation of the action (\ref{1}) w.r.t $ g_{df} $ yields the $f(R,T)$ field equation as,\\
\begin{eqnarray}\label{2}
f_{R}(R,T) R_{df} - \frac{1}{2}f(R,T)g_{df} + (g_{df} \Box - \nabla_{d}\nabla_{f})f_{R}(R,T) =
\nonumber \\
 8 \pi T_{df} - f_{T}(R,T)T_{df} -  f_{T}(R,T)n_{df}.
\end{eqnarray}\\
Here $ f_{R}(R,T) = \frac{\partial f(R,T)}{\partial R}, f_{T}(R,T) = \frac{\partial f(R,T)}{\partial T}, \Box = \partial _{d} (\sqrt{-g}g^{df}\partial_{f})/\sqrt{-g} , R_{df} = $ Ricci tensor , $ \nabla_{d} = $ covariant derivative compatible with $ g_{df} $, and \\
\begin{equation}\label{3}
n_{df} = g^{\nu\zeta} \frac{\delta T_{\nu\zeta}}{\delta g^{df}}.
\end{equation}\\
The stress-energy tensor takes the form of\\
\begin{equation}\label{4}
T_{df} = g_{df} \hat{L}_{M} - \frac{2\delta \hat{L}_{M}}{\delta g^{df}}.
\end{equation}\\
Taking the covariant derivative of equation (\ref{2}), we get\\
\begin{eqnarray}\label{5}
\nabla^{d} T_{df} &=& \frac{ f_{T}(R,T)}{8 \pi -  f_{T}(R,T)}[(T_{df} + n_{df})\nabla^{d} \ln  [f_{T}(R,T)]
\nonumber \\
&& - \nabla^{f} n_{df} - \frac{1}{2}g_{df}\nabla^{d} T ].
\end{eqnarray}\\
Evidently, the stress-energy tensor in $ f(R,T) $ theory isn't conserved, unlike GR. For our model, the stellar interior is composed of perfect fluid. Thus,\\
\begin{equation}\label{6}
T_{df} = (\rho + p )u_{d}u_{f} - pg_{df}
\end{equation}\\
for $ u^{d}u_{d} = 1 $ and $ u^{d}\nabla_{f}u_{d} = 0 $. Setting $ \hat{L}_{M} = -p $ \cite{Q20},\\
\begin{equation}\label{7}
n_{df} = -2T_{df} - pg_{df}.
\end{equation}\\
In accordance with \cite{T1}, we consider $ f(R,T) $ as $ f(R,T) = R + 2 m(T) $, where $ m(T) $ is the function of the stress energy tensor's trace. We take $ m(T) = \eta T $ to specify the modified gravitational theory with $ \eta $ as the coupling constant. This form has proven to be instrumental to derive various cosmological, astrophysical solutions \cite{Q19,T2,T7}. Hence, from equation (\ref{2}) we find,\\
\begin{equation}\label{8}
\textsf{G}_{df} = 8 \pi T_{df} + \eta Tg_{df} + 2\eta (T_{df} + pg_{df})
\end{equation}\\
where $ \textsf{G}_{df} $ stands for the Einstein tensor. When $ f(R,T) = R $ or $ \eta = 0 $, we retrieve the GR results. Plugging the above functional form in equation (\ref{5}), we get\\
\begin{equation}\label{9}
\nabla^{d} T_{df} = \frac{-2 \eta}{(8 \pi + 2\eta)}[\nabla^{d}(pg_{df}) + \frac{1}{2}g_{df}\nabla^{d}T].
\end{equation}\\
Substituting $ \eta = 0 $ verifies the energy momentum tensor conservation akin to GR. Due to its benefits in addressing various cosmological and astrophysical studies, this gravity theory is very much favoured \cite{Q19,Q20}. The line element describing the interior spacetime of the star with spherical symmetry is\\
\begin{equation}\label{10}
ds^2=e^{i(r)}dt^2 - e^{j(r)}dr^2 - r^2(d\theta^2+\sin^2\theta d\phi^2)
\end{equation}\\
with $ i(r) $ and $ j(r) $ as radial functions. The set of field equations we can have corresponding to equation (\ref{10}) are as\\
\begin{equation}\label{11}
d(r)[-1 + e^{j} + j'r] = 8 \pi \rho + \eta(3 \rho - p )
\end{equation}\\
\begin{equation}\label{12}
d(r)[-1 + e^{j} -i'r] = -8 \pi p + \eta( \rho - 3p )
\end{equation}\\
\begin{equation}\label{13}
d(r)[\frac{r}{2}(j'-i') - \frac{(2i'' + i'^{2} - i'j')}{4} r^{2} ] = -8 \pi p + \eta( \rho - 3p )
\end{equation}\\
with $ d(r) = e^{-j}r^{-2} $, where prime signifies derivative w.r.t $ r $. The energy momentum tensor's non-conservation in this theory leads to\\
\begin{equation}\label{14}
p' + \frac{i'}{2}(\rho + p) + \frac{\eta}{(8 \pi + 2 \eta)}(p' - \rho') = 0.
\end{equation}\\
For $ \textsl{m} $ as the gravitational mass with radius $ r $, equation (\ref{11}) implies\\
\begin{equation}\label{15}
e^{-j} = 1 - \frac{2 \textsl{m}}{r} - \eta(\rho - \frac{p}{3}) r^{2}.
\end{equation}\\
From the combination of the equations (\ref{14}) and (\ref{15}), we have\\
\begin{equation}\label{16}
p' = -(\rho + p) \frac{[ 4 \pi r p - \frac{\eta(\rho - 3 p)}{2}r ] + [ \frac{\textsl{m}}{r^{2}} + \frac{\eta}{2}(\rho - \frac{p}{r})r]}{[1-\frac{2\textsl{m}}{r} - \eta(\rho - \frac{p}{3})r^{2}][1 + \frac{\eta(1 - \frac{dp}{d\rho})}{2(4 \pi + \eta)}]}.
\end{equation}\\
Assuming vacuum spacetime in the exterior surrounding it, our solution for the interior must be consistent with the exterior Schwarzschild solution produced as\\
\begin{equation}\label{17}
ds^2 = \big(1 - \frac{2\emph{M}}{r}\big){dt}^2 - \big(1 - \frac{2\emph{M}}{r}\big)^{-1}dr^2 - r^2(d\theta^2+\sin^2\theta d\phi^2).
\end{equation}\\
Here $ \emph{M} $ denotes the total mass. For $ r = R $,  at the boundary\\
\begin{equation}\label{18}
e^{-i(R)} = e^{j(R)} = 1 - \frac{2\emph{M}}{R}.
\end{equation}\\
Employing the DP approach, we now analyze the effects by studying its implications. We adopt the form of DP modified metric as \cite{D1,D3},\\
\begin{equation}\label{19}
e^{j(r)} = \frac{7 + 14 Fr^{2} + 7Fr^{4}}{7 - 10Fr^{2} - F^{2}r^{4}}
\end{equation}\\
where this metric $ e^{j(r)} $ is advantageous due to its freedom from all types of singularities throughout. This also leads to a physically viable energy profile, decreasing and finite everywhere inside. Upon the substitution of equation (\ref{19}) into the field equations (\ref{11}-\ref{13}), we get\\
\begin{eqnarray}\label{20}
\Big(\frac{7 - 10Fr^{2} - F^{2}r^{4}}{7 + 14 Fr^{2} + 7Fr^{4}}\Big)\Big(- \frac{(28F + 28Fr^{2})(7 - 10Fr^{2} - F^{2}r^{4})}{(7 + 14 Fr^{2} + 7Fr^{4})^{2}}
\nonumber \\
\frac{-20F - 4F^{2}r^{2}}{7 + 14 Fr^{2} + 7Fr^{4}} - \frac{1}{r^{2}}\Big) + \frac{1}{r^{2}} = 8 \pi \rho + \eta(3 \rho - p )
\end{eqnarray}\\
\begin{equation}\label{21}
\Big(\frac{7 - 10Fr^{2} - F^{2}r^{4}}{7 + 14 Fr^{2} + 7Fr^{4}}\Big)\Big(\frac{-i'}{r} - \frac{1}{r^{2}} \Big) + \frac{1}{r^{2}} = -8 \pi p + \eta( \rho - 3p )
\end{equation}\\
\begin{eqnarray}\label{22}
\frac{1}{r^{2}}\Big(\frac{7 - 10Fr^{2} - F^{2}r^{4}}{7 + 14 Fr^{2} + 7Fr^{4}}\Big)\Big(\frac{r}{2}\big(\frac{-20Fr - 4F^{2}r^{3}}{7 + 14 Fr^{2} + 7Fr^{4}} - \frac{(28Fr + 28Fr^{3})(7 - 10Fr^{2} - F^{2}r^{4})}{(7 + 14 Fr^{2} + 7Fr^{4})^{2}} -i'\big)
\nonumber \\
- \Big(2i'' + i'^{2} - i'\big(\frac{-20F - 4F^{2}r^{2}}{7 + 14 Fr^{2} + 7Fr^{4}} - \frac{1}{r^{2}} - \frac{(28F + 28Fr^{2})(7 - 10Fr^{2} - F^{2}r^{4})}{(7 + 14 Fr^{2} + 7Fr^{4})^{2}}\big)\Big)\frac{r^{2}}{4}\Big) = -8 \pi p + \eta( \rho - 3p ).
\end{eqnarray}\\

\section{BEC stellar paradigm}\label{sec3}

\hspace{0.7cm}Compact stellar objects composed of BECs are referred as BEC stars \cite{R13,R16}. Previously, GR has been used to investigate the BEC star properties. BEC stars may contribute to the elusive DM phenomena \cite{R22}. BEC stars at non-zero temperature and chemical potential have been explored, as we can find in the literature \cite{R,S1}.Properties of BEC stars in modified gravity are studied via BEC matter descriptions. The relativistic EoS for BEC stars derived from the Gross-Pitaevskii framework is represented as \cite{R16}\\
\begin{equation}\label{23}
p(\rho) = U \rho^{2} = \frac{l \rho^{2}}{2\textsf{M}^{2}}
\end{equation}\\
where $ l = \frac{ 4 \pi \alpha \hbar^{2}}{\textsf{M}} $ characterizes the interaction strength. We set the condensate particle mass to be $ \textsf{M} = 2 m $ with $ m = 1.675 \times 10^{-24} $g ( nucleon mass ). This reflects the possibility of nucleons pairing into bosonic states. Given $ \alpha = 1 fm $ scattering length, the interaction strength becomes $ 4.17 \times 10^{-43} $ g $ cm^{5} \textbf{s}^{2} $ \cite{S}. Equation (\ref{18}) along with the boundary conditions at $ r = R $, we determine the constant $ F $ as 
\begin{table}[h!]
\centering
\caption{Determination of the parameter value }
\begin{tabular}{||p{5.5cm}|p{2.9cm}|p{2.0cm}|p{2.0cm}||}
\hline\hline
\hspace{1cm}$ Compact \hspace{0.2cm} star $ & \hspace{0.8cm}$ M(M_{\bigodot}) $ & \hspace{0.8cm}$ R $(km) & \hspace{0.8cm}$ F $ \\
\hline\hline
$\hspace{1.1cm} PSR-B0943 + 10 $ & \hspace{0.7cm}$ 0.2 $ & $ 2.6 $ & $ 0.0000149 $\\[1pt]
\hline
$\hspace{1.1cm} CEN X-3 $ & \hspace{0.7cm}$ 1.49 $ & $ 4.178 $ & $ 0.0000696 $\\[1pt]
\hline
$\hspace{1.1cm} SMC X-4 $ & \hspace{0.7cm}$ 1.29 $ & $ 8.831 $ & $ 0.0000283 $\\[1pt]
\hline
$\hspace{1.1cm} HER X-1 $ & \hspace{0.7cm}$ 0.85 $ & $ 8.1 $ & $ 0.0000203 $\\[1pt]
\hline
$\hspace{1.1cm} 4U 1538-58 $ & \hspace{0.7cm}$ 0.87 $ & $ 7.8666 $ & $ 0.0000214 $\\[1pt]
\hline\hline
\end{tabular}
\label{Tab:1}
\end{table}\\
\begin{figure}[h!]
\centering
\includegraphics[scale=0.5]{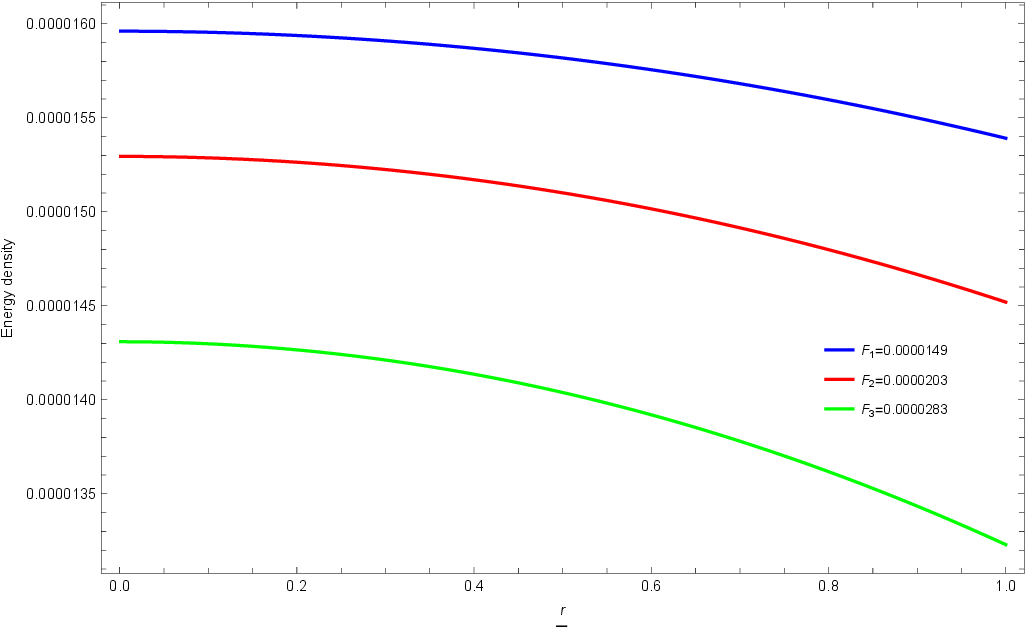}
\caption{Behavior of the density profile w.r.t $ \frac{r}{R} $ for $ PSR-B0943 + 10, HER X-1, SMC X-4 $ and $ \eta = 0.2 $ }\label{1}
\end{figure}\\
\begin{figure}[h!]
\centering
\includegraphics[scale=0.5]{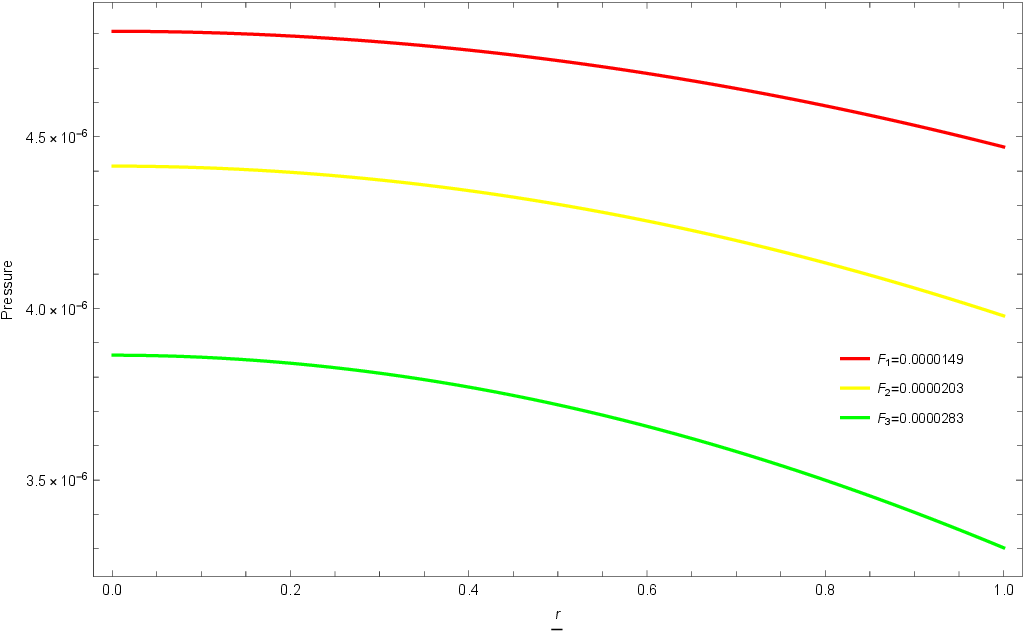}
\caption{Behavior of the pressure w.r.t $ \frac{r}{R} $ for $ PSR-B0943 + 10, HER X-1, SMC X-4 $ and $ \eta = 0.2 $ }\label{2}
\end{figure}\\
Numerical derivation yields the density and the pressure outcomes for our stellar model in the interior. Figures (\ref{1}) and (\ref{2}) demonstrate pressure and density positivity, peaking at the center and decreasing radially outward for $ 0 < r < R $. This shows the model's physical soundness \cite{Q19}.\\

\subsection{Energy conditions}

\hspace{0.5cm}The stellar system requires the following energy inequalities to satisfy. Verifying them in $ f(R,T) $ gravity's tensor shows the proposed BEC star model is physically viable. It provides as a crucial benchmark for our model's validation. They are given by\\
\begin{equation}\label{24}
Null \hspace{0.2cm} Energy \hspace{0.2cm} Condition ( NEC ): \rho + p \geq 0
\end{equation}\\
\begin{equation}\label{25}
Weak \hspace{0.2cm} Energy \hspace{0.2cm} Condition ( WEC ):\rho \geq 0, \hspace{0.3cm} \rho + p \geq 0
\end{equation}\\
\begin{equation}\label{26}
Strong \hspace{0.2cm} Energy \hspace{0.2cm} Condition ( SEC ):  \rho + 3p \geq 0
\end{equation}\\
\begin{equation}\label{27}
Dominant \hspace{0.2cm} Energy \hspace{0.2cm} Condition ( DEC ): \rho - p \geq 0.
\end{equation}\\
We can clearly observe from the figures (\ref{3}-\ref{5}) that all the energy conditions are met throughout, and consistently distributed within the interior. Satisfaction of these conditions enhances confidence in stability and eliminates unphysical effects.\\
\begin{figure}[h!]
\centering
\includegraphics[scale=0.5]{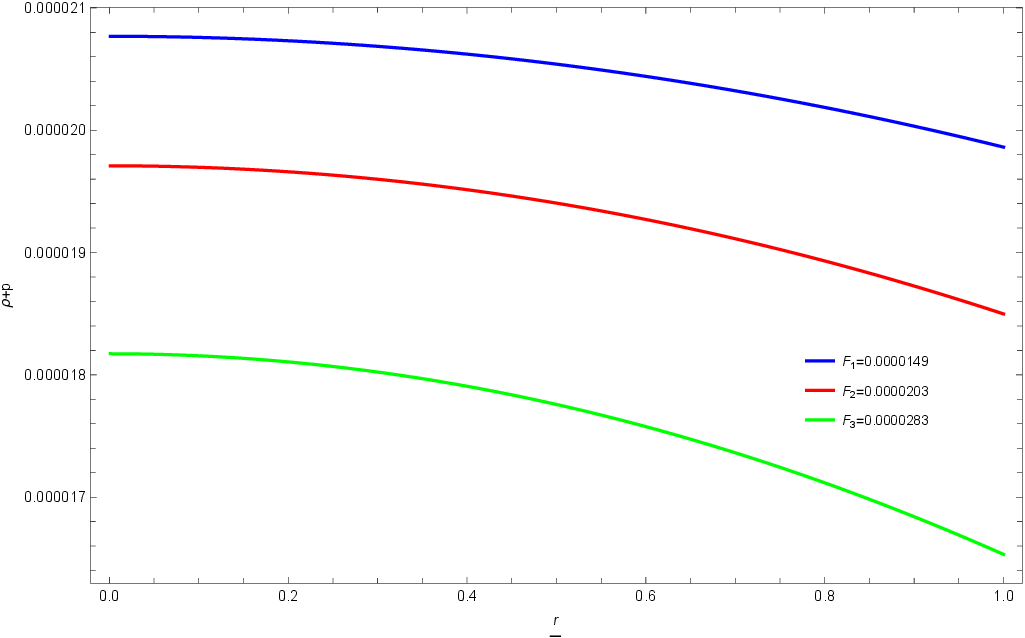}
\caption{Plot of $ \rho + p $ vs $ \frac{r}{R} $ for different parameter values with $ \eta = 0.2 $ }\label{3}
\end{figure}\\
\begin{figure}[h!]
\centering
\includegraphics[scale=0.5]{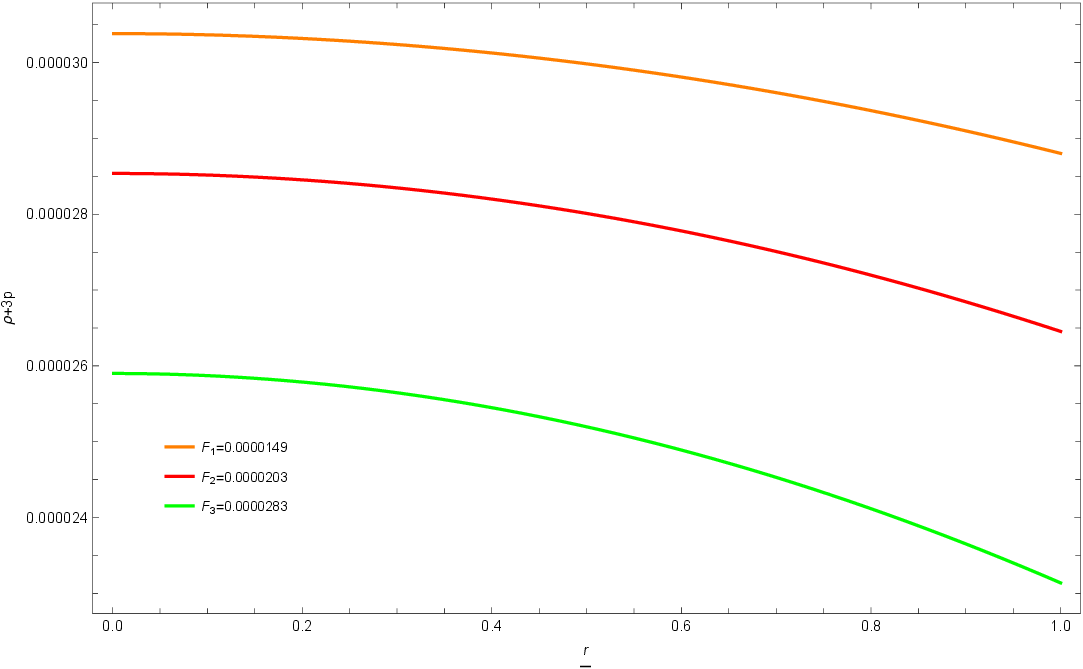}
\caption{Plot of $ \rho + 3p $ vs $ \frac{r}{R} $ for different parameter values with $ \eta = 0.2 $ }\label{4}
\end{figure}\\
\begin{figure}[h!]
\centering
\includegraphics[scale=0.5]{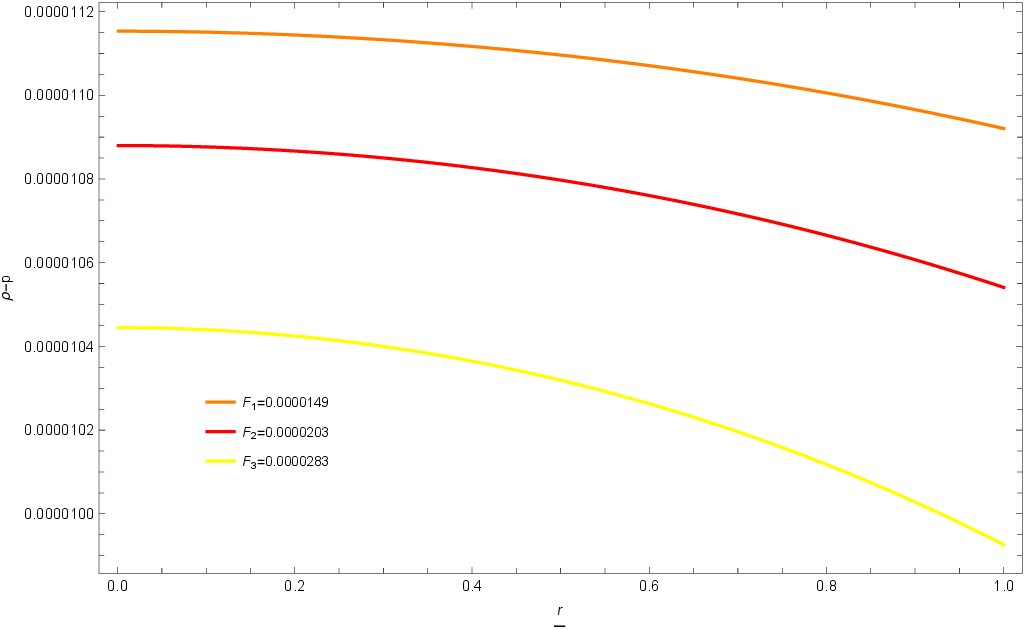}
\caption{Plot of $ \rho - p $ vs $ \frac{r}{R} $ for different parameter values with $ \eta = 0.2 $  }\label{5}
\end{figure}\\

\subsection{EoS parameter}

\hspace{0.5cm}Here our analysis centers on the EoS parameter $ \omega $ with density, and pressure gradients in the interior. The EoS parameter is determined by the relation of\\
\begin{equation}\label{28}
\omega = \frac{p}{\rho}.
\end{equation}
\begin{figure}[h!]
\centering
\includegraphics[scale=0.5]{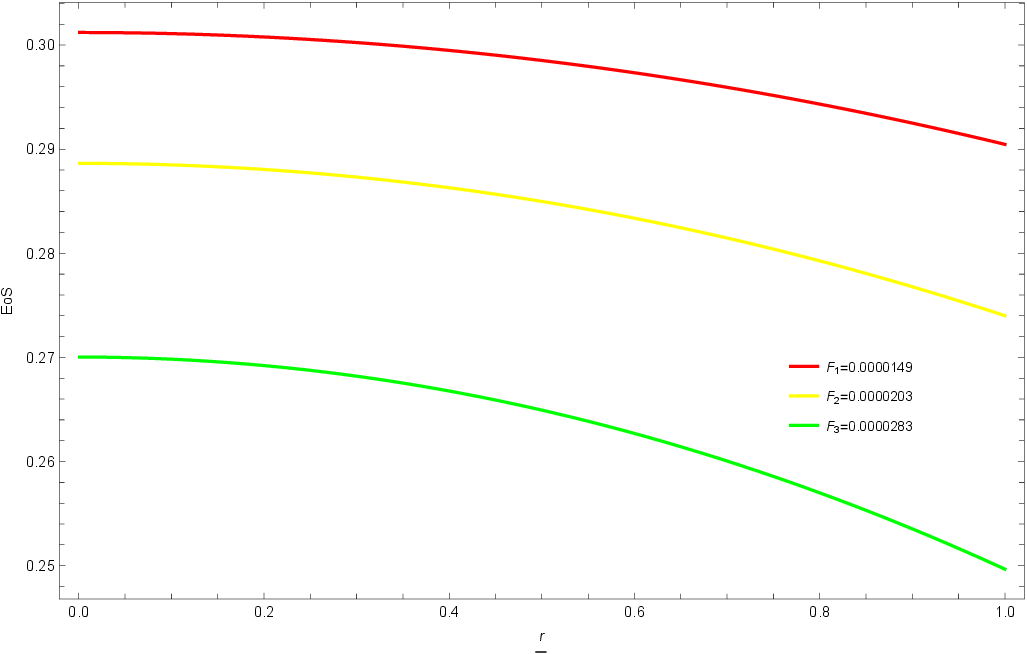}
\caption{Nature of $ \omega $ vs $ \frac{r}{R} $ for different parameter values with $ \eta = 0.2 $  }\label{6}
\end{figure}\\
\begin{figure}[h!]
\centering
\includegraphics[scale=0.5]{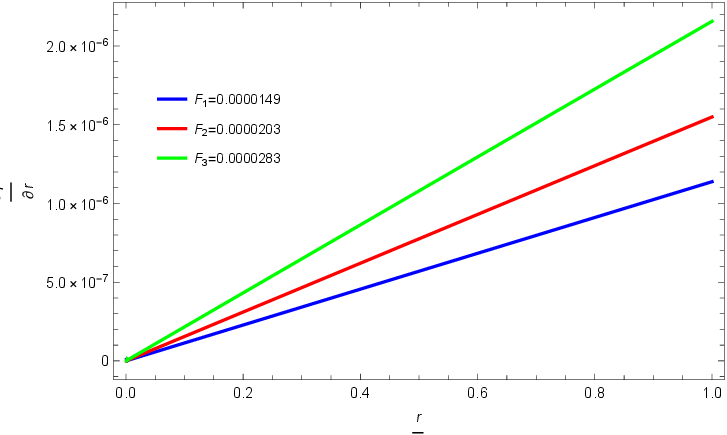}
\caption{Nature of $ \partial_{r} \rho $ vs $ \frac{r}{R} $ for different parameter values with $ \eta = 0.2 $  }\label{7}
\end{figure}\\ 
\begin{figure}[h!]
\centering
\includegraphics[scale=0.5]{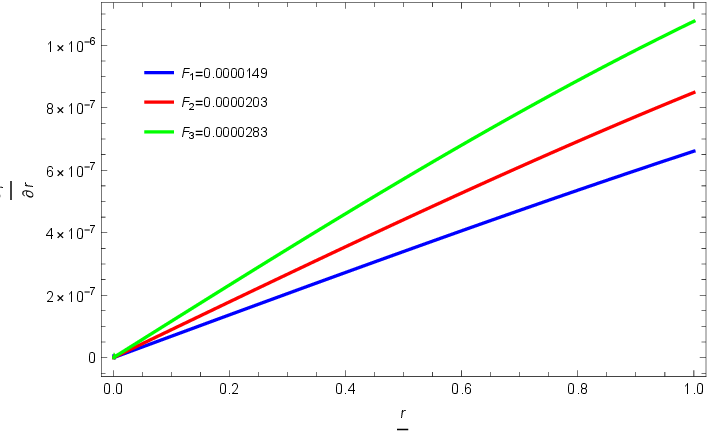}
\caption{Nature of $ \partial_{r} p $ vs $ \frac{r}{R} $ for different parameter values with $ \eta = 0.2 $  }\label{8}
\end{figure}\\
The graphical representation of the solution is given in figure (\ref{6}). The results confirm with the predicted range of $ [0,1] $. We further analyze the gradients of the energy-momentum tensor components, depicted in figures (\ref{7}) and(\ref{8}). Evidently, the gradients are positive and their behavior are observed as expected \cite{BQ}.\\

\section{Validity conditions for stellar structure}\label{sec4}

\hspace{0.7cm} We graphically illustrated and assessed through our model's effectiveness and stability in this section. The following subsections discuss about sound velocity, adiabetic index and surface redshift.\\

\subsection{Sound velocity}

\hspace{0.5cm}Stability demands the sound velocity to be less than light velocity, viz., $ 0 < \textsc{V}^{2} < 1 $, which ensures that the compact object obey casuality, a basic necessity. This would establish a stable and realistic stellar model. The velocity of sound is given by,\\
\begin{equation}\label{29}
\textsc{V}^{2} = \frac{dp}{d\rho} = \frac{dp}{dr}\frac{dr}{d\rho}.
\end{equation}.
\begin{figure}[h!]
\centering
\includegraphics[scale=0.5]{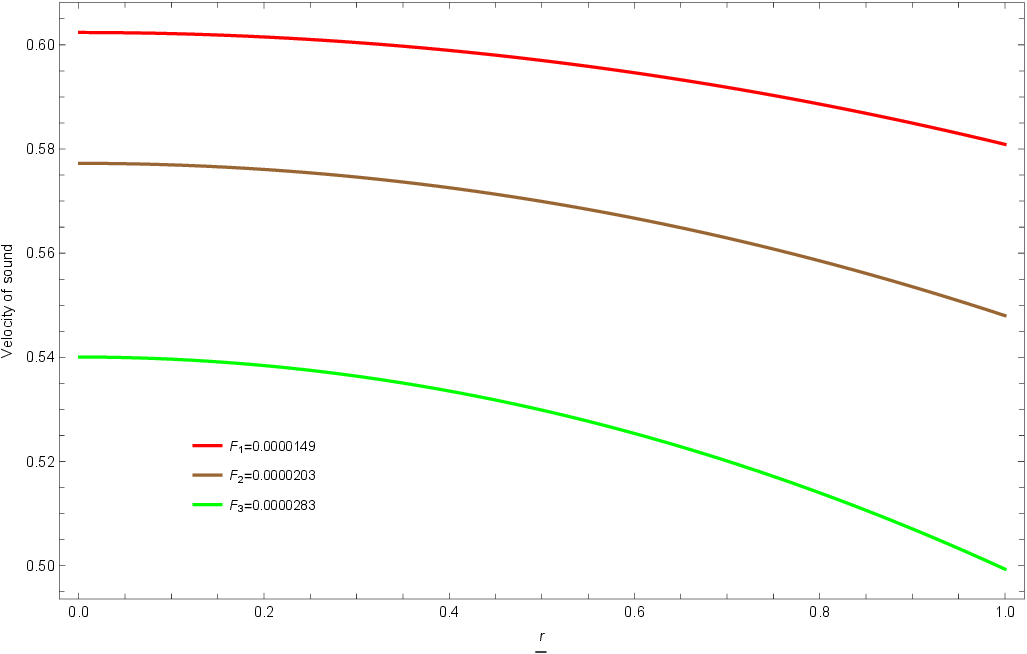}
\caption{ $ \textsc{V}^{2} $ is plotted w.r.t $ \frac{r}{R} $ for $ F = 0.0000149, F = 0.0000203, F = 0.0000283 $ with $ \eta = 0.2 $  }\label{9}
\end{figure}\\
Figure (\ref{9}) shows that the parameter $ \textsc{V}^{2} $ is decreasing radially outward and also remaining well below the speed of light. Thus we can rely on our hypothesized BEC star model within $ f(R,T) $ gravity obtained via DP metric function.\\

\subsection{Adiabetic index}

\hspace{0.5cm}The adiabetic index, crucial for the stability analysis. It describes the sensitivity of pressure to density variations. The threshold often maintains a value greater than $ \frac{4}{3} $ with slight shifts. It is defined as\\
\begin{equation}\label{30}
\acute{\Gamma} = (\frac{\rho + p}{p})\frac{dp}{d\rho}
\end{equation}\\
\begin{figure}[h!]
\centering
\includegraphics[scale=0.5]{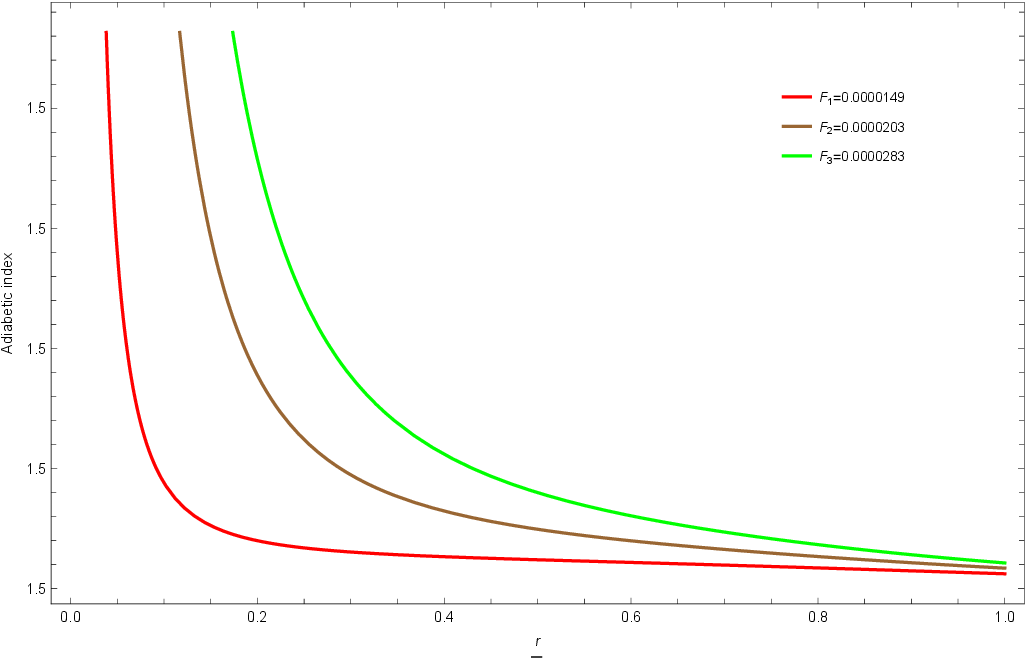}
\caption{ $ \acute{\Gamma} $ is plotted w.r.t $ \frac{r}{R} $ for $ F = 0.0000149, F = 0.0000203, F = 0.0000283 $ with $ \eta = 0.2 $  }\label{10}
\end{figure}\\
where we have$ \frac{dp}{d\rho} $ as $ \frac{dp}{d\rho} = \frac{dp}{dr}\frac{dr}{d\rho} $. The adiabetic index (figure (\ref{10})) supports the stability (i.e. $ \acute{\Gamma} > \frac{4}{3} $) for our BEC model in modified $ f(R,T) $ gravity. The model exhibits adiabetic stability internally, but smaller $ F $ , that is for the values $ F \ll 0 $ might disrupt their stability and greater parameter value destines to be more stable within this gravitational theory.\\

\subsection{Surface redshift}

\hspace{0.5cm} To strengthen the stability and formulation of our proposed model we have also taken the help of surface redshift analysis. If the redshift stays within the upper limit of $ 2 $ with positive backdrop throughout the interior it implies a stable configuration. The model's redshift $ \texttt{Z}_{S} $ is formulated as\\
\begin{equation}\label{31}
\texttt{Z}_{S} = e^{-\frac{i}{2}} - 1
\end{equation}
\begin{figure}[h!]
\centering
\includegraphics[scale=0.5]{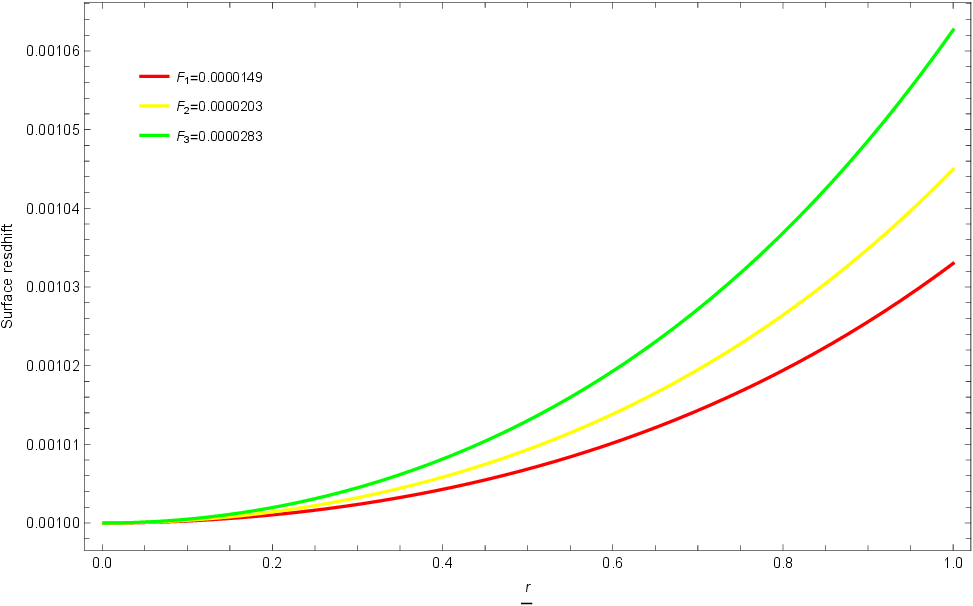}
\caption{ $ \texttt{Z}_{S} $ is plotted w.r.t $ \frac{r}{R} $ for $ F = 0.0000149, F = 0.0000203, F = 0.0000283 $ with $ \eta = 0.2 $  }\label{11}
\end{figure}\\
where positive values are expected within the structures. Figure (\ref{11}) illustrates the surface redshift's range i.e. within $ 0 $ and $ 2 $ affirming our model's matter distribution's validity. Thus a stable, casual compact model is suggested which supports to the gravity-matter coupling constraints.\\

\section{Discussion and conclusion}\label{sec5}

\hspace{0.7cm}This study examines the finite temperature BEC stars in $ f(R,T) $ gravity adopting the DP approach. We have also broadened our model to study analytical solutions of diverse stellar frameworks. Our work employs the finite temperature BEC EoS based on the Gross-Pitaevskii equation. Our numerical results indicate isotropic pressure, decreasing to zero at the surface and also the outward decreasing density ( figures (\ref{1}) and (\ref{2}) ), where we have used the obtained values of $ F $ from the table, matching the two spacetimes at $ r= R $. We now highlight our findings for the three models under study with the coupling constant $ \eta = 0.2 $.\\
\begin{itemize}
\item{\textbf{Energy conditions}} : Our proposed model meets the energy constraints everywhere inside the star for varying parameter values as pictured in figures (\ref{3}-\ref{5}).\\
\item{\textbf{EoS parameter}} : The next aspect of our study is the EoS parameter. The DP model in $ f(R,T) $ gravity yields physically viable BEC stars with $ \omega $ ranging from 0 to 1.\\
\item{\textbf{Gradients of the energy-momentum tensor components}} : Our investigation included examining $ \frac{\partial \rho}{\partial r} $ and $ \frac{\partial p}{\partial r} $. The gradients decrease steadily, significantly less than 1. Additionally, it guarantees the velocity of sound would not surpass that of light. The corresponding plots are displayed in figures (\ref{7}) and (\ref{8}).\\
\item{\textbf{Velocity of sound}} : Numerical outcomes confirm that the parameter $ \textsc{V}^{2} $'s values fall within the acceptable range of 0 and 1 for our proposed framework as visualized in figure (\ref{9}).\\
\item{\textbf{Adiabetic index}} : The adiabetic index analysis assesses the stability of compact objects. Figure (\ref{10}) indicates adiabetic stability for higher $ F $ values, instability for $ F \ll 0 $.\\
\item{\textbf{Redshift}} : A redshift limit of 2 is expected for physically viable compact stars. Our DP model for $ f(R,T) $ gravity adhered to this constraint. Our findings are visually represented in figure (\ref{11}).\\
\end{itemize}
Thus our model within this gravity incorporating specific DP metric function confirms tha configuration's precision and admissibility in representing astrophysical structures. It is worthwhile to explore such BEC stellar models in different alternative gravitational frameworks utilizing a variety of metric potentials.\\

\end{document}